%% file: 00_paper.tex
\documentclass{INTERSPEECH2023}

\interspeechcameraready

\usepackage{amsmath,graphicx}
\usepackage{tabularx}
\usepackage{booktabs}
\usepackage{multirow}
\usepackage{cleveref}
\usepackage{flushend}
\usepackage{placeins}
\usepackage{url}
\usepackage{xspace}
\usepackage{siunitx}
\usepackage[shortcuts]{extdash}

\newcommand{\fb}[0]{FluencyBank\xspace}
\newcommand{\sep}[0]{SEP\=/28k\xspace}
\newcommand{\sepE}[0]{SEP\=/28k\=/E\xspace}
\newcommand{\ksof}[0]{KSoF\xspace}

\newcommand{\ie}{\textit{i}.\textit{e}., }

\title{A Stutter Seldom Comes Alone -- Cross-Corpus Stuttering Detection as a Multi-label Problem}

\name{Sebastian P. Bayerl$^{1}$,
Dominik Wagner$^{1}$, 
Ilja Baumann$^{1}$, 
Florian Hönig$^{2}$,
Tobias Bocklet$^{1,3}$,\\ 
Elmar Nöth$^{4}$,
Korbinian Riedhammer$^{1}$}

\address{
$^{1}$ Technische Hochschule Nürnberg Georg Simon Ohm, Germany \\
$^{2}$KST Institut GmbH, $^{3}$Intel Labs\\
$^{4}$ Friedrich-Alexander-Universität Erlangen-Nürnberg, Germany
}
\email{sebastian.bayerl@ieee.org}

\begin{document}

\maketitle
 
\begin{abstract}

Most stuttering detection and classification research has viewed stuttering as a multi-class classification problem or a binary detection task for each dysfluency type; however, this does not match the nature of stuttering, in which one dysfluency seldom comes alone but rather co-occurs with others.
This paper explores multi-language and cross-corpus end-to-end stuttering detection as a multi-label problem using a modified wav2vec 2.0 system with an attention-based classification head and multi-task learning.
We evaluate the method using combinations of three datasets containing English and German stuttered speech, one containing speech modified by fluency shaping.
The experimental results and an error analysis show that multi-label stuttering detection systems trained on cross-corpus and multi-language data achieve competitive results but performance on samples with multiple labels stays below overall detection results.

\end{abstract}

\input{01_introduction}

\input{03_data}

\input{02_method}
\input{04_experiments}

\input{05_discussion}

\input{06_conclusion}

\FloatBarrier
\newpage

\vfill
\pagebreak

\bibliographystyle{IEEEtran}
\bibliography{zotero}

\end{document}

%% file: 01_introduction.tex
\section{Introduction}\label{sec:intro}
Dysfluency refers to abnormalities of fluency, which includes, but is not limited to, stuttering \cite{wingate_FluencyDisfluencyDysfluency_1984}.
Stuttering is a complex fluency disorder distinguished by its core symptoms, which include repetitions of words, syllables, and sounds, as well as prolongations and blocks while speaking \cite{lickley_DisfluencyTypicalStuttered_2017}.
Therapeutic options such as speech modification have been proposed to alleviate symptoms and improve fluency.
Stuttering can impair a person's ability to communicate, which extends to voice technology, necessitating the detection of atypical speech and the application of custom models.

Recent work on machine learning for stuttering focused on stuttering detection \cite{bayerl_DetectingDysfluenciesStuttering_2022a, harvill_FramelevelStutterDetection_2022,kourkounakis_DetectingMultipleSpeech_2020} and stuttering classification \cite{grosz_Wav2vec2basedParalinguisticSystems_2022,sheikh_EndtoEndSelfSupervisedLearning_2022,montacie_AudioFeaturesWav2Vec_2022,you_MaskedModelingbasedAudio_2022}. 
Grosz et al. used pre-trained German wav2vec 2.0 (W2V2) models and combined them with other classifiers in an ensemble approach \cite{grosz_Wav2vec2basedParalinguisticSystems_2022}.
Montacie et al. also used W2V2 features, treating them like low-level descriptors, and computed several functionals on the features, similar to the openSMILE approach \cite{montacie_AudioFeaturesWav2Vec_2022,eyben_OpensmileMunichVersatile_2010}.

In \cite{kourkounakis_DetectingMultipleSpeech_2020}, the authors describe an approach employing long short-term memory (LSTM) networks with a residual neural network (ResNet) backend to detect stuttering in the UCLASS corpus \cite{howell_UniversityCollegeLondon_2009,kourkounakis_DetectingMultipleSpeech_2020}.
Lea et al. applied multi-task learning with LSTMs in their dysfluency detection approach and treated stuttering detection as a multi-label problem but did not evaluate the multi-label performance \cite{lea_SEP28kDatasetStuttering_2021}.
They could show that dysfluency detection systems trained on one dataset generalized to another, given enough training data \cite{lea_SEP28kDatasetStuttering_2021}. 
Recent work by Bayerl et al. showed that W2V2 feature extractors could be fine-tuned for stuttering detection using English stuttering data \cite{bayerl_DetectingDysfluenciesStuttering_2022a}.
The respective features were transferable from English to German for all dysfluency types except word repetitions.
Both contributions did not explore the effect of multi-language and cross-dataset training or evaluation on their detection systems.  

The \sep dataset, and the relabeled \fb dataset, are fairly new resources.
Unfortunately, the authors did not publish the data partitioning, making it hard to reproduce and compare results to their baseline systems \cite{lea_SEP28kDatasetStuttering_2021}.
Several researchers evaluated their methods but did not disclose their exact splits either, or filtered out examples that substantially impact evaluation results.
Sheikh et al. removed a class by combining word- and sound repetitions and filtered out clips with an additional non-stuttering label, e.g., natural pause or background music \cite{lea_SEP28kDatasetStuttering_2021}, i.e., removing difficult examples.
They randomly split the remaining data into a train, development, and test set (80/10/10\%), irrespective of the speakers \cite{sheikh_RobustStutteringDetection_2022}. 
Other work filtered out all non-unanimously labeled clips from the training and test data, aiming for easy samples \cite{mohapatra_SpeechDisfluencyDetection_2022}.
The work by \cite{jouaiti_DysfluencyClassificationStuttered_2022} left out the block class and applied random splitting irrespective of the speaker, which can lead to overly optimistic results, as a study on the influence of dataset partitioning could show  \cite{bayerl_InfluenceDatasetPartitioning_2022}.
As a possible solution, the authors extended the dataset by adding semi-auto-generated speaker labels and non-speaker overlapping splits of \sep, including results for baseline experiments \cite{bayerl_InfluenceDatasetPartitioning_2022}.

Apart from problems with reproducibility, most works ignore that one stutter seldom comes alone, \ie multiple types of dysfluencies co-occur. 
In {SEP\=/28k\=/Extended (\sepE)}, \fb, and \ksof, about 23\%, 29\%, and 17\% of clips were labeled with more than one dysfluency relevant label type -- strong evidence that stuttering detection and classification are in fact multi-label problems.

Our contributions are 1) reproducible multi-label stuttering detection and classification using an end-to-end (E2E) system based on W2V2 and evaluated on three datasets, one containing modified speech (fluency shaping); 2) experimental evidence for the generalizability and feasibility of multi-language and cross-dataset training of multi-label dysfluency detection systems; 3) a detailed analysis regarding specialized binary vs. multi-label dysfluency classification that shows that classification errors happen more often on samples with multiple dysfluency labels.

%% file: 03_data.tex
\vspace{-2mm}
\section{Data}\label{sc:data}
\vspace{-2mm}
In our experiments, we use three corpora containing 3-second long clips with stuttered speech; SEP-28k-Extended, FluencyBank, and the Kassel State of Fluency (\ksof) dataset \cite{bayerl_InfluenceDatasetPartitioning_2022,lea_SEP28kDatasetStuttering_2021,bayerl_KSoFKasselState_2022}. 
The SEP-28k-E contains about 28.000 English stuttered speech extracted from podcasts and is based on the SEP-28k corpus, extending it with speaker labels and a speaker-exclusive Train-Dev-Test split.\footnote{Online: \protect\url{https://tinyurl.com/yck9fmfv}}

The \sep corpus contains an additional 4144 English clips extracted from the interview part of the adults who stutter dataset of the \fb corpus \cite{bernsteinratner_FluencyBankNew_2018} that were labeled using the same protocol.
For evaluation purposes, we use the split defined by \cite{bayerl_DetectingDysfluenciesStuttering_2022a}.\footnote{\protect Online: \url{https://tinyurl.com/24vm6dec}}
All three datasets were labeled similarly, containing no dysfluencies or one or more types of dysfluencies; blocks, prolongations, sound repetitions, word repetitions, and interjections.
\ksof is a German dataset containing 5597 segments extracted segments from stuttering therapy recordings. 
In addition to the dysfluency labels in \sepE and \fb, the clips were also labeled with speech modifications, marking a clip as containing a person using fluency shaping.
Fluency shaping is a technique persons who stutter (PWS) learn in stuttering therapy to help them overcome their stuttering \cite{bayerl_KSoFKasselState_2022}.
The exact label distribution of the datasets can be found in the literature \cite{lea_SEP28kDatasetStuttering_2021,bayerl_KSoFKasselState_2022,bayerl_InfluenceDatasetPartitioning_2022}.
All datasets have ambiguously labeled segments, unlike \sepE and \fb; KSoF does not have samples labeled as belonging to the no dysfluencies class while at the same time being marked as containing any of the five-dysfluency labels.

To evaluate the effect of multi-lingual and multi-dataset training, we define three combinations of datasets.
ALL-EN combines \fb and \sepE, Multi-Lingual-Small (M-Ling-S) combines \ksof and \fb, and Multi-Lingual (M-Ling) consists of \ksof, \fb, and \sepE. 
Systems training with X-Ling-S, M-Ling, and \ksof are trained to solve a seven-class multi-label classification problem (Modified (Mod), Blocks (Bl), Interjections (Int), Prolongation (Pro), Sound Repetitions (Snd), Word Repetitions (Wd), No Dysfluencies (No-Df)).
While the \fb, \sepE, and ALL-EN splits are trained to detect six classes (Bl, Int, Pro, Snd, Wd, No-Df).

%% file: 02_method.tex
\vspace{-2mm}
\section{Method}\label{sec:method}

\subsection{wav2vec 2.0}\label{ss:w2v2}
\vspace{-2mm}
In our experiments, we use large W2V2 models, as \cite{grosz_Wav2vec2basedParalinguisticSystems_2022} have shown that large W2V2 models with more parameters outperform the base models with fewer parameters in the related multi-class stuttering classification task \cite{grosz_Wav2vec2basedParalinguisticSystems_2022}.
The weights for the models used in the experiments were pre-trained for English or German ASR, respectively, and are available in the referenced source code. 
The large W2V2 model consists of a convolutional feature extractor at the beginning of the model that takes in the waveform, followed by 24 transformer encoder blocks.  
The model encodes 20ms of audio into 1024-dimensional feature vectors after each transformer block, yielding 24 x $t$ x 1024-dimensional hidden representations, with $t=\frac{time\ s}{20ms}$ \cite{baevski_Wav2vecFrameworkSelfSupervised_2020}. 
W2V2 features perform well in dysfluency detection \cite{bayerl_DetectingDysfluenciesStuttering_2022a} and other speech tasks, such as automatic speech recognition (ASR), and mispronunciation detection \cite{baevski_Wav2vecFrameworkSelfSupervised_2020,xu_ExploreWav2vecMispronunciation_2021}.

\begin{figure}[!ht]
    \centering
    \includegraphics[width=0.9\linewidth]{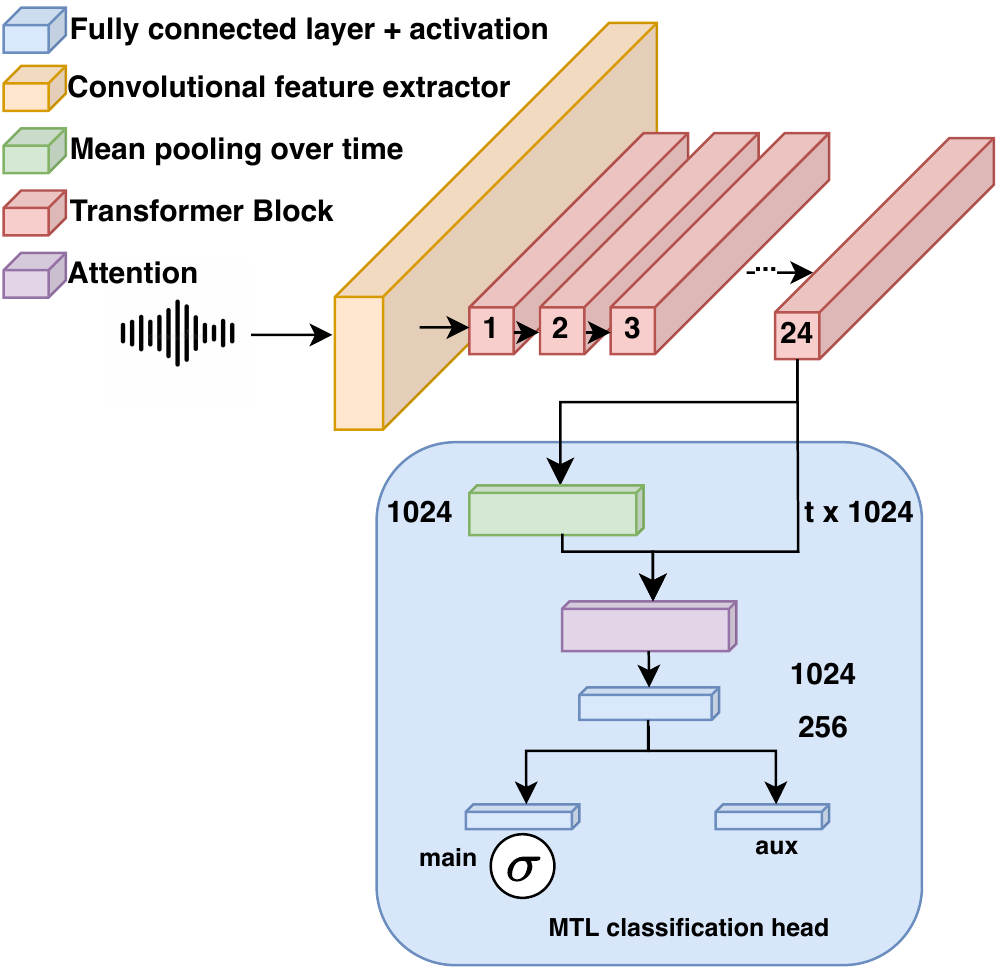}
    \caption{Schematic overview of the wav2vec 2.0 model with modified attention-based multi-task (MTL) classification head.   
    }
  \label{fig:architecture}
  \vspace{-5mm}
\end{figure}

The model was adapted from the standard W2V2 sequence classification implementation \cite{wolf_TransformersStateoftheArtNatural_2020} and differs mainly in three aspects:
Instead of using only a single output branch, we use an alternative output branch for multi-task learning using a second output branch. 
Instead of using mean pooling along the time dimension, a global scaled dot-product attention mechanism takes the mean of all feature vectors along the time dimension as the query input $Q$, with keys $K$ and values $V$ being the mean over time for each dimension of the hidden states from the last hidden state \cite{vaswani_AttentionAllYou_2017}.   
This is done to better account for the different temporal contexts needed to detect each of the dysfluency classes than just using mean-pooling over time can do.
The third change is a second output branch used for multi-task learning, analogous to the one described in \cite{bayerl_DetectingDysfluenciesStuttering_2022a}. 
A sigmoid function ($\sigma$) is applied to the main task outputs predicting the probability of each class for an audio clip.
All models reported have an auxiliary branch with two outputs and a softmax activation function. 
A schematic representation of the components and changes to the default classification head is depicted in \Cref{fig:architecture}. 

\vspace{-2mm}
\subsection{Loss}\label{ss:mtl}

Previous studies have shown the usefulness of multi-task learning (MTL) when training dysfluency detection systems, using either an artificial `any' label, indicating the presence of any dysfluency, or gender classification. \cite{lea_SEP28kDatasetStuttering_2021, bayerl_DetectingDysfluenciesStuttering_2022a,sheikh_RobustStutteringDetection_2022}.  

In our experiments, we use a combination of weighted Binary Cross Entropy (BCE) and Focal Loss (FL) \cite{lin_FocalLossDense_2020}. 
FL is an extension of the BCE loss using the $\alpha$ and $\gamma$ parameters to put special emphasis on minority classes to handle class imbalance. 

\begin{equation}\label{eq:focal_loss}
    \mathbf{FL}(p_t) = -\alpha (1 - p_t)^\gamma \log{(p_t)}.
\end{equation}

FL can be used equivalently to BCE loss for multi-label problems by calculating the loss for each class using the output of a given output neuron. 
In our experiments, the overall FL is calculated 
by computing the mean loss value for each class as shown in \cref{eq:focal_loss_multi}.

\begin{equation}\label{eq:focal_loss_multi}
    \vspace{-2mm}
    \mathbf{FL_{multi}} = \frac{1}{n}\sum^n_1{\mathbf{FL_n}(p_t)}
\end{equation}

The final multi-task loss is computed by combining the BCE loss for the auxiliary task ($L_{\text{aux}}$) and FL for the main task ($L_{\text{main}}$) 
as a sum weighted by $w_{main}$, as shown in \cref{eq:mtlloss}.

\begin{equation}\label{eq:mtlloss}
    \vspace{-2mm}
    \mathbf{L}_{\text{MTL}} = w_{\text{main}} L_{\text{main}} + (1-w_{\text{main}}) L_{\text{aux}}
\end{equation}

%% file: 04_experiments.tex
\section{Experiments}\label{sec:experiments}

Our experiments provide insights into dysfluency detection as a multi-label problem and the influence of training data quantity and composition across datasets and languages. 
Preliminary experiments included modifying the classification head of the W2V2 model by adding an attention mechanism for pooling with a trainable token parameter, mean and statistical pooling as implemented in \cite{wolf_TransformersStateoftheArtNatural_2020}, or a classification token-based mechanism as used by BERT \cite{devlin_BERTPretrainingDeep_2019}, which led to slightly worse results than the employed attention-based classification head. 

All experiments were performed using publicly available models of large pre-trained W2V2 feature extractors. 
The experiments using M-Ling, M-Ling-S, \fb, \sepE, and ALL-EN training partitions were based on a model that was 
fine-tuned for ASR on the LibriSpeech corpus (ASR LARGE (EN)) \cite{baevski_Wav2vecFrameworkSelfSupervised_2020, panayotov_LibrispeechASRCorpus_2015}.
The experiments using only \ksof data for training were based on a fine-tuned model for German ASR using the Common Voice dataset (ASR LARGE (DE)).\footnote{Code available: 
\protect\url{https://tinyurl.com/mr4h3ybk}}
To assess the effects of the multi-language and cross-corpus training, we also performed fine-tuning experiments (exp.~10 -- 15), utilizing the weights obtained by training experiment 1 (cf. \Cref{tab:results_data}). 

All systems were trained for up to 20 epochs using the adamW optimizer, a warm-up phase of 10\% of total training steps and a patience of five epochs, choosing the best model based on the development loss \cite{loshchilov_DecoupledWeightDecay_2019}.
During training, the convolutional feature extractor at the input of the model was frozen. 
The best hyper-parameters for training were determined using Optuna \cite{akiba_OptunaNextgenerationHyperparameter_2019} with the tree-structured Parzen estimator (TPE) sampling strategy. 
The optimization algorithm could choose from an initial learning rate, $lr \in \{\num{3e-5}, \num{3e-5}\}$, and batch-sizes $bs \in \{8, 32, 64, 128, 256\}$.  
The main loss weight $w_{main}$ and the FL parameters $\alpha$ and $\gamma$ were determined from $w_{main} \in \{0.85,... 0.95\}$,  $\gamma \in \{1, 2, 3\}$, and $\alpha \in \{0.1, 0.2,  \ldots, 0.9 \}$. 
The auxiliary task was chosen from either gender classification or an artificial ``any'' label, indicating the presence of any dysfluency in a segment, for experiments using only a single dataset and the ALL-EN combination. 
Experiments using M-Ling-S and M-Ling data also used language-id as their potential auxiliary task. 
The best-performing parameters across all experiments were $lr=\num{3e-5}$, $bs=8$, $\alpha=0.7$, and $\gamma=3$.
The best performing auxiliary task for experiments using only the \sepE and ALL-EN data was detecting any dysfluency with $w_{\text{main}}=0.9$. 
For \ksof and \fb, gender classification led to the best overall results with $w_{\text{main}}=0.92$. 
Training using the M-Ling and M-Ling-S data profited most from the language-id  task with $w_{\text{main}}=0.92$.

%% file: 05_discussion.tex
\section{Results and Discussion}\label{sec:results}

This section describes the overall and specific results from \Cref{tab:results_data} and \Cref{tab:xling_results_data}.
All systems are evaluated on their training data. 
Their cross-dataset and cross-language performance is tested by evaluating the models on the corpora that were not used for training.
We report F1 scores for all dysfluency types and modifications.
The result N/A indicates cases where precision and recall are zero per definition, as there are no labeled clips in the respective partition, i.e., F1 is undefined.
The results were balanced between \textit{precision} and \textit{recall}.
\Cref{tab:multi_label_results} considers multi-label classification metrics for each dataset.

\input{05_results_table_per_dataset}

\Cref{tab:results_data} shows the effect of training data quantity.
The system trained on the larger \sepE dataset generalizes well to \fb, improving or matching results reported in the literature with expert binary dysfluency detection systems (Baseline systems), unlike the system trained using only \fb which generalizes poorly to \sepE and \ksof. 
The multi-label systems trained on \sepE or \ksof, improve or match the baseline results for all dysfluency classes except \textbf{Wd} for \ksof and \textbf{Bl} for \sepE.

The fine-tuning experiments using \ksof data mostly help to improve the results in the cross-dataset setting substantially but otherwise fail to improve results over the systems based on the ASR LARGE (DE) model by a large margin. 
Results for \fb improve substantially over the system trained using \fb and the ASR LARGE (EN) model.

A comparison of \Cref{tab:xling_results_data} and \Cref{tab:results_data} reveals that the models trained on the multi-language and multi-dataset data (M-Ling-S, M-Ling) achieve the overall best results, except for \textbf{Int} on \fb and \textbf{Wd} on \ksof (exp.~19--24). 
The models achieve the best per-dysfluency performance in most cases, even though the models trained on the multi-language data have to account for the additional \textbf{Mod} class, showing their potential to generalize across languages.

The model trained using \ksof data achieved the highest F1 for \textbf{Mod} (exp.~9), which is matched by the model trained using the M-Ling-S data.
Performance decreased slightly in the model trained with M-Ling data (exp.~21, 24).
Evaluating the recognition of \textbf{Mod} with the multi-lingual models shows no confusion for \textbf{Mod} to any English segments and vice-versa.  
This might be due to the language or the distinctiveness of the pattern, but also, different recording conditions may cause these results.

\input{05_cross_lingual_results}

To evaluate the multi-label performance, we use the exact match ratio (EMR), the partial match ratio (PMR), recall, and the Hamming loss (HL). 
The EMR is a pessimistic evaluation metric, extending accuracy to the multi-label case \cite{sorower_LiteratureSurveyAlgorithms_2010}, ignoring the notion of partially correct samples. 
The PMR is the ratio of samples for which at least one label was correctly recalled to the total number of samples. 
The HL specifies the fraction of incorrectly classified labels; it considers both the prediction error and the missing error normalized by the total number of classes and examples \cite{sorower_LiteratureSurveyAlgorithms_2010}.

The results in \Cref{tab:multi_label_results} and the following analysis consider the overall best-performing model trained on the M-Ling data. 
The EMR results for all three datasets are around 0.5. 
Only looking at the segments that have more than one label drops these results to around 0.2, indicating that many classification errors occur in segments that have multiple labels. 
The increased HL indicates that the model makes about twice as many mistakes on segments with multiple labels.  
In our error analysis of the multi-label results, we, therefore, looked at the most common combinations of segments labeled with two types of dysfluencies for each dataset and picked three label combinations for discussion.
To be included in the analysis, there had to be at least 50 segments for each combination in the dataset. 
Results are contained in \Cref{tab:combi_mistakes}

\textbf{Int} and \textbf{Wd} are the most frequent combination of two dysfluencies in \sepE and \fb.
The results show a higher EMR in the multi-label case for this label combination than for the overall dataset, with a very high partial match rate. 
\textbf{Ints} are generally recognized well across all systems, and the model tends to favor its prediction.
In the partial matches, the prediction of \textbf{Int} dominates in cases where the model only predicts one of the labels correctly.
The predictions for the combination of \textbf{Int} and \textbf{Snd} are similar, with higher EMR values for the label combination than the overall dataset and a high PMR value.
Still, \textbf{Ints} are recalled more often than \textbf{Snd}. 
Again, the model favors predicting only \textbf{Ints}, which seems to hinder the prediction of \textbf{Snd}.  
These co-occurrences and the favored prediction of \textbf{Ints} have implications for the evaluation of binary detection systems, where it is possible that the model rather learns to recognize the \textbf{Ints} in the segments rather than the target dysfluency or is distracted by the most dominant pattern. 
This can influence results and has to be accounted for during training time if the pattern co-occurs often enough.  

The combination of \textbf{Pro} and \textbf{Bl} has a zero EMR for SEP-28k-E. 
The model fails to predict the combination completely and also has a very low EMR for \ksof.
The models have difficulties detecting \textbf{Bl} on their own (see Bl results in \Cref{tab:xling_results_data}) and even more so in the multi-label case.  
Assessing \textbf{Bl} using only acoustics is a hard task, even for clinicians that usually rely on signs of physical tension and grasping for air when assessing blocks. 
This is reflected by low inter-rater reliability (IRR)(Fleiss $\kappa$, 0.25 SEP-28k, 0.37 \ksof) for \textbf{Bl} reported for the datasets \cite{lea_SEP28kDatasetStuttering_2021, bayerl_KSoFKasselState_2022}. 
Only the results for \ksof manage to improve over previous systems (Baseline) and manage to recall 42\% of \textbf{Bl} in the multi-label case (see \Cref{tab:combi_mistakes}). 
We hypothesize that 
this is due to \ksof consisting of therapy recordings that include PWS who, on average, have more pronounced symptoms, which is supported by the higher $\kappa$.

The analysis of multi-label errors shows the necessity for measures to deal with multi-label dysfluency data during training time. 
Possible measures might include modified loss functions and the inclusion of prior knowledge about common co-occurring dysfluencies.

\input{05_multi_label_results}

\input{03_data_combis}

%% file: 05_results_table_per_dataset.tex
\begin{table}[!ht]
    \centering
    \vspace{-2mm}
    \caption{
        Multi-label dysfluency detection results (F1-score) using multi-label E2E systems.
        (\textbf{Mod} = Modified Speech, \textbf{Bl} = Block, \textbf{Int} = Interjection, \textbf{Pro} = Prolongation, \textbf{Snd} = Sound repetition, \textbf{Wd} = Word repetition).
        Section headers indicate the training data, followed by the W2V2 weights used.
    }
    \scalebox{0.90}{
        \begin{tabular}{c|c|c|c|c|c|c|c}
            \toprule
            \# & \textbf{Test}                                             & \textbf{Mod} & \textbf{Bl} & \textbf{Int} & \textbf{Pro} & \textbf{Snd} & \textbf{Wd} \\
            \midrule
            \multicolumn{8}{c}{\textbf{Baseline Systems}}                                                                                                                  \\
            \midrule
            -  & \sepE \cite{bayerl_InfluenceDatasetPartitioning_2022}     & -            & 0.33        & 0.68         & 0.46         & 0.39         & 0.51        \\
            -  & FBANK \cite{bayerl_DetectingDysfluenciesStuttering_2022a} & -            & 0.33        & 0.84         & 0.60         & 0.60         & 0.43        \\
            -  & \ksof \cite{bayerl_DetectingDysfluenciesStuttering_2022a} & 0.76         & 0.54        & 0.74         & 0.53         & 0.47         & 0.19        \\
            \midrule
            \multicolumn{8}{c}{\textbf{\sepE (ASR LARGE EN)}}                                                                                                      \\
            \midrule
            1  & \textbf{\sepE}                                            & -          & 0.16        & \textbf{0.77}         & \textbf{0.51}         & \textbf{0.50}         & \textbf{0.62}        \\
            2  & FBANK                                                     & -          & 0.17        & \textbf{0.82}        & 0.60        & \textbf{0.63}         & \textbf{0.47}       \\
            3  & \ksof                                                     & -          & 0.10        & 0.55         & 0.44         & 0.35         & \textbf{0.23}       \\
            \midrule
            \multicolumn{8}{c}{\textbf{\fb (ASR LARGE EN)}}                                                                                                        \\
            \midrule
            4  & \sepE                                                     & -          & 0.00        & 0.65         & 0.41         & 0.42         & 0.16        \\
            5  & \textbf{FBANK}                                            & -          & 0.00        & 0.73         & 0.44         & 0.55         & 0.25        \\
            6  & \ksof                                                     & -          & 0.00        & 0.46         & 0.35         & 0.38         & 0.06        \\
            \midrule
            \multicolumn{8}{c}{\textbf{\ksof (ASR LARGE DE)}}                                                                                                      \\
            \midrule
            7  & \sepE                                                     & N/A          & 0.27        & 0.55         & 0.40         & 0.42         & 0.11        \\
            8  & FBANK                                                     & N/A          & 0.27        & 0.60         & 0.56         & 0.58         & 0.12        \\
            9  & \textbf{\ksof}                                            & \textbf{0.76}        & \textbf{0.60}       & \textbf{0.88}         & \textbf{0.57}         & 0.48         & 0.18        \\

            \midrule
            \multicolumn{8}{c}{\textbf{\fb (Stutter LARGE EN)}}                                                                                                    \\
            \midrule
            10 & \textit{\sepE}                                            & NaN          & 0.22        & 0.72         & 0.49         & 0.46         & 0.64        \\
            11 & \textbf{FBANK}                                            & NaN          & 0.24        & 0.79         & \textbf{0.61}         & 0.60         & 0.51        \\
            12 & \ksof                                                     & NaN          & 0.30        & 0.59         & 0.42         & 0.49         & 0.19        \\
            \midrule
            \multicolumn{8}{c}{\textbf{KSoF (Stutter LARGE EN)}}                                                                                                   \\
            \midrule
            13 & \textit{\sepE}                                            & NaN          & \textbf{0.30}       & 0.67         & 0.39         & 0.42         & 0.25        \\
            14 & FBANK                                                     & NaN          & \textbf{0.29}        & 0.66         & 0.51         & 0.57         & 0.11        \\
            15 & \textbf{\ksof}                                            & 0.65         & 0.49        & 0.79         & 0.53         & \textbf{0.51}         & 0.11        \\
            \bottomrule
        \end{tabular}
    }
    \vspace{-3mm}
    \label{tab:results_data}
\end{table}

%% file: 05_cross_lingual_results.tex
\begin{table}[!ht]
    \centering
    \vspace{-2mm}
    \caption{
        Cross-Dataset and multi-Lingual dysfluency detection results (F1-score) using E2E multi-label systems.
        (\textbf{Mod} = Modified Speech, \textbf{Bl} = Block, \textbf{Int} = Interjection, \textbf{Pro} = Prolongation, \textbf{Snd} = Sound repetition, \textbf{Wd} = Word repetition).
        Headers indicate the training data, followed by the W2V2 weights used.
    }
    \scalebox{0.90}{

        \begin{tabular}{c|c|c|c|c|c|c|c}

            \toprule
            \# & \textbf{Test} & \textbf{Mod} & \textbf{Bl} & \textbf{Int} & \textbf{Pro} & \textbf{Snd} & \textbf{Wd} \\
            \midrule
            \multicolumn{8}{c}{\textbf{ALL-EN (ASR LARGE EN)}}                                                         \\
            \midrule
            16 & \sepE         & -          & 0.22        & 0.74         & 0.51         & 0.49         & 0.63        \\
            17 & FBANK         & -          & 0.23        & \textbf{0.80}         & 0.59         & 0.63         & 0.51        \\
            18 & \ksof         & -          & 0.35        & 0.65         & 0.37         & 0.47         & \textbf{0.18}        \\
            \midrule
            \multicolumn{8}{c}{\textbf{Multi-Lingual-Small (ASR LARGE EN)}}                                            \\
            \midrule
            19 & \sepE         & N/A          & 0.25        & 0.70         & 0.50         & 0.45         & 0.26        \\
            20 & FBANK         & N/A          & 0.09        & 0.75         & 0.56         & 0.61         & 0.40        \\
            21 & \ksof         & \textbf{0.76}        & 0.56        & \textbf{0.90}         & 0.58         & \textbf{0.54}         & 0.11        \\
            \midrule
            \multicolumn{8}{c}{\textbf{Multi-Lingual (ASR LARGE EN)}}                                                  \\
            \midrule
            22 & \sepE         & N/A          & \textbf{0.32}        & \textbf{0.77}         & \textbf{0.53}         & \textbf{0.53}         & \textbf{0.64}        \\
            23 & FBANK         & N/A          & \textbf{0.36}        & 0.79         & \textbf{0.62}         & \textbf{0.64}         & \textbf{0.52}       \\
            24 & \ksof         & 0.75         & \textbf{0.64}        & 0.85         & \textbf{0.60}         & 0.48         & 0.14        \\

            \bottomrule
        \end{tabular}
    }
    \label{tab:xling_results_data}
    \vspace{-4mm}

\end{table}

%% file: 05_multi_label_results.tex
\begin{table}[!ht]
    \centering
    \vspace{-2mm}
    \caption{Hamming Loss (HL) and Equal Match Ratio (EMR) for
        the test set of each dataset (Test) and segments from the test set that were
        labeled with more than one dysfluency (Test-Multi)}
    \begin{tabular}{l|cc|cc}
                         & \multicolumn{2}{c}{\textbf{Test}} & \multicolumn{2}{c}{\textbf{Test Multi}}                              \\

        \toprule
        \textbf{Dataset} & \textbf{HL}                       & \textbf{EMR}                            & \textbf{HL} & \textbf{EMR} \\
        \midrule
        \sepE            & 0.13                              & 0.54                                    & 0.23        & 0.2          \\
        FBANK            & 0.11                              & 0.53                                    & 0.21        & 0.23         \\
        \ksof            & 0.11                              & 0.49                                    & 0.18        & 0.21         \\
    \end{tabular} \label{tab:multi_label_results}
    \vspace{-3mm}
\end{table}

%% file: 03_data_combis.tex
\begin{table}[!hp]
    \centering
\vspace{-2mm}    
    \caption{Equal match ratio, partial match ratio (PMR), and recall (Re) for the first (L1) and second label (L2).}
    \scalebox{0.94}{
    \begin{tabular}{l|c|c|c|c|c}
        L1 \& L2                        &Dataset & EMR     & PMR & Re L1   & Re L2    \\
        \toprule                      
        
        \textbf{Int} \& \textbf{Wd}  &\sepE &0.44 & 0.95 &            0.79 &           0.60    \\
\textbf{Int} \& \textbf{Wd}  &\fb &  0.39 & 0.97 & 0.88 & 0.48  \\
        \midrule
\textbf{Int} \& \textbf{Snd} & \sepE & 0.36 & 0.90 &    0.74 &  0.51  \\
\textbf{Int} \& \textbf{Snd} & \fb&  0.33 & 0.87 &  0.80 &  0.40  \\
\textbf{Int} \& \textbf{Snd} & \ksof & 0.29 & 0.93 & 0.79 &  0.43  \\
\midrule
\textbf{Pro} \& \textbf{Bl} & \sepE & 0.00 & 0.76 & 0.69 &  0.07  \\
\textbf{Pro} \& \textbf{Bl} & \ksof &0.16 & 0.84 & 0.58 &  0.42  \\
    \end{tabular}\label{tab:combi_mistakes}
    }
\end{table}

%% file: 06_conclusion.tex
\vspace{-2mm}
\section{Conclusion}\label{sec:conclusion}

This paper introduced a multi-label W2V2-based end-2-end stuttering detection and classification system. 
Leveraging multi-language training data from multiple datasets, previous results achieved by binary per-dysfluency classification systems could, for the most part, be improved on each of the \fb, \sepE, and \ksof datasets, even though multi-label detection is a more complex task. 

Furthermore, results emphasize the difficulty of co-occurring dysfluencies when training binary-detection systems. 
For multi-label systems, these patterns should be accounted for during training time to improve detection results and prevent errors on test samples with multiple labels. 
The experimental results show that the quantity and diversity of training data are still the most important factors for creating more robust dysfluency detection systems, even if using large pre-trained models such as W2V2. 
This is even more true in the case of multi-label stuttering detection, where there are only a few samples for each dysfluency combination.

In future work, we will therefore explore prior-knowledge-infused handling of segments with co-occurring dysfluencies as well as data augmentation techniques to increase or improve the respective classification performance.